\documentclass[a4paper,aps,pre,reprint,superscriptaddress]{revtex4-1}

\usepackage[latin1]{inputenc}    
\usepackage{amsmath}           
\usepackage{amsfonts}          
\usepackage{amssymb}           
\usepackage{lmodern,dsfont}	
\usepackage{color}
\usepackage{graphicx}          
\usepackage[T1]{fontenc}       

\newcommand{\motSizeI}{1.8ex}
\newcommand{\motSizeII}{2.1ex}
\newcommand{\motSizeIII}{1.8ex}
\newcommand{\motSizeIV}{0.9ex}

\newcommand{\ie}{\textit{i.e.~}}

\newcommand{\motI}{
 {\mathchoice
  {\includegraphics[height=\motSizeI]{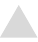}}
  {\includegraphics[height=\motSizeII]{triadConfig_1.pdf}}
  {\includegraphics[height=\motSizeIII]{triadConfig_1.pdf}}
  {\includegraphics[height=\motSizeIV]{triadConfig_1.pdf}}
 }
}
\newcommand{\motII}{
 {\mathchoice
  {\includegraphics[height=\motSizeI]{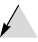}}
  {\includegraphics[height=\motSizeII]{triadConfig_2.pdf}}
  {\includegraphics[height=\motSizeIII]{triadConfig_2.pdf}}
  {\includegraphics[height=\motSizeIV]{triadConfig_2.pdf}}
 }
}
\newcommand{\motIII}{
 {\mathchoice
  {\includegraphics[height=\motSizeI]{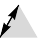}}
  {\includegraphics[height=\motSizeII]{triadConfig_3.pdf}}
  {\includegraphics[height=\motSizeIII]{triadConfig_3.pdf}}
  {\includegraphics[height=\motSizeIV]{triadConfig_3.pdf}}
 }
}
\newcommand{\motIV}{
 {\mathchoice
  {\includegraphics[height=\motSizeI]{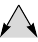}}
  {\includegraphics[height=\motSizeII]{triadConfig_4.pdf}}
  {\includegraphics[height=\motSizeIII]{triadConfig_4.pdf}}
  {\includegraphics[height=\motSizeIV]{triadConfig_4.pdf}}
 }
}
\newcommand{\motV}{
 {\mathchoice
  {\includegraphics[height=\motSizeI]{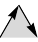}}
  {\includegraphics[height=\motSizeII]{triadConfig_5.pdf}}
  {\includegraphics[height=\motSizeIII]{triadConfig_5.pdf}}
  {\includegraphics[height=\motSizeIV]{triadConfig_5.pdf}}
 }
}
\newcommand{\motVI}{
 {\mathchoice
  {\includegraphics[height=\motSizeI]{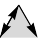}}
  {\includegraphics[height=\motSizeII]{triadConfig_6.pdf}}
  {\includegraphics[height=\motSizeIII]{triadConfig_6.pdf}}
  {\includegraphics[height=\motSizeIV]{triadConfig_6.pdf}}
 }
}
\newcommand{\motVII}{
 {\mathchoice
  {\includegraphics[height=\motSizeI]{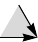}}
  {\includegraphics[height=\motSizeII]{triadConfig_7.pdf}}
  {\includegraphics[height=\motSizeIII]{triadConfig_7.pdf}}
  {\includegraphics[height=\motSizeIV]{triadConfig_7.pdf}}
 }
}
\newcommand{\motVIII}{
 {\mathchoice
  {\includegraphics[height=\motSizeI]{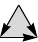}}
  {\includegraphics[height=\motSizeII]{triadConfig_8.pdf}}
  {\includegraphics[height=\motSizeIII]{triadConfig_8.pdf}}
  {\includegraphics[height=\motSizeIV]{triadConfig_8.pdf}}
 }
}
\newcommand{\motIX}{
 {\mathchoice
  {\includegraphics[height=\motSizeI]{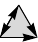}}
  {\includegraphics[height=\motSizeII]{triadConfig_9.pdf}}
  {\includegraphics[height=\motSizeIII]{triadConfig_9.pdf}}
  {\includegraphics[height=\motSizeIV]{triadConfig_9.pdf}}
 }
}
\newcommand{\motX}{
 {\mathchoice
  {\includegraphics[height=\motSizeI]{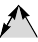}}
  {\includegraphics[height=\motSizeII]{triadConfig_10.pdf}}
  {\includegraphics[height=\motSizeIII]{triadConfig_10.pdf}}
  {\includegraphics[height=\motSizeIV]{triadConfig_10.pdf}}
 }
}
\newcommand{\motXI}{
 {\mathchoice
  {\includegraphics[height=\motSizeI]{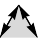}}
  {\includegraphics[height=\motSizeII]{triadConfig_11.pdf}}
  {\includegraphics[height=\motSizeIII]{triadConfig_11.pdf}}
  {\includegraphics[height=\motSizeIV]{triadConfig_11.pdf}}
 }
}
\newcommand{\motXII}{
 {\mathchoice
  {\includegraphics[height=\motSizeI]{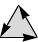}}
  {\includegraphics[height=\motSizeII]{triadConfig_12.pdf}}
  {\includegraphics[height=\motSizeIII]{triadConfig_12.pdf}}
  {\includegraphics[height=\motSizeIV]{triadConfig_12.pdf}}
 }
}
\newcommand{\motXIII}{
 {\mathchoice
  {\includegraphics[height=\motSizeI]{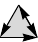}}
  {\includegraphics[height=\motSizeII]{triadConfig_13.pdf}}
  {\includegraphics[height=\motSizeIII]{triadConfig_13.pdf}}
  {\includegraphics[height=\motSizeIV]{triadConfig_13.pdf}}
 }
}
\newcommand{\motXIV}{
 {\mathchoice
  {\includegraphics[height=\motSizeI]{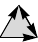}}
  {\includegraphics[height=\motSizeII]{triadConfig_14.pdf}}
  {\includegraphics[height=\motSizeIII]{triadConfig_14.pdf}}
  {\includegraphics[height=\motSizeIV]{triadConfig_14.pdf}}
 }
}
\newcommand{\motXV}{
 {\mathchoice
  {\includegraphics[height=\motSizeI]{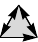}}
  {\includegraphics[height=\motSizeII]{triadConfig_15.pdf}}
  {\includegraphics[height=\motSizeIII]{triadConfig_15.pdf}}
  {\includegraphics[height=\motSizeIV]{triadConfig_15.pdf}}
 }
}
\newcommand{\motXVI}{
 {\mathchoice
  {\includegraphics[height=\motSizeI]{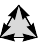}}
  {\includegraphics[height=\motSizeII]{triadConfig_16.pdf}}
  {\includegraphics[height=\motSizeIII]{triadConfig_16.pdf}}
  {\includegraphics[height=\motSizeIV]{triadConfig_16.pdf}}
 }
}

\begin{document}
\title{Motifs in Triadic Random Graphs based on Steiner Triple Systems}
\author{Marco Winkler}
\email{mwinkler@physik.uni-wuerzburg.de}
\author{J\"{o}rg Reichardt}
\email{reichardt@physik.uni-wuerzburg.de}

\affiliation{Institute for Theoretical Physics, University of W\"{u}rzburg, Am Hubland, 97074 W\"{u}rzburg, Germany}
	
\date{\today}
\begin{abstract}
Conventionally, pairwise relationships between nodes are considered to be the fundamental building blocks of complex networks. However, over the last decade the overabundance of certain sub-network patterns, so called motifs, has attracted high attention. It has been hypothesized, these motifs, instead of links, serve as the building blocks of network structures.

Although the relation between a network's topology and the general properties of the system, such as its function, its robustness against perturbations, or its efficiency in spreading information is the central theme of network science, there is still a lack of sound generative models needed for testing the functional role of subgraph motifs. Our work aims to overcome this limitation.

We employ the framework of exponential random graphs (ERGMs) to define novel models based on triadic substructures. The fact that only a small portion of triads can actually be set independently poses a challenge for the formulation of such models. To overcome this obstacle we use Steiner Triple Systems (STS). These are partitions of sets of nodes into pair-disjoint triads, which thus can be specified independently. Combining the concepts of ERGMs and STS, we suggest novel generative models capable of generating ensembles of networks with non-trivial triadic Z-score profiles. Further, we discover inevitable correlations between the abundance of triad patterns, which occur solely for statistical reasons and need to be taken into account when discussing the functional implications of motif statistics. Moreover, we calculate the degree distributions of our triadic random graphs analytically.
\end{abstract}

\maketitle


\section{Introduction}
The topological structure of interactions among the constituents of complex many particle systems is intimately linked to system function and global system properties. The study of complex networks aims to elucidate this link between structure and function. 

Motivated by the stark contrast between topological features found in real-world data and expectations based on the assumption of purely random link formation \cite{Erdos1959,Erdos1960}, two main threads of research can be identified.

The first thread is aimed predominantly at explaining the network formation process, \ie identifying the forces shaping a network. A particularly productive approach has been the development of network growth models following the publication of Barab\'asi and Albert \cite{Barabasi1999} to explain non-Poissonian degree distributions. See \cite{Albert2002,krapivsky2003,Dorogovtsev2003} and the references therein for a review. Growth models generally take the agreement between a particular feature in real-world data with networks resulting from a particular model as evidence for a particular aspect of a growth process, such as preferential attachment.

The second thread of research focusses on explaining the influence certain topological features may have on global system properties such as the robustness against perturbations or the stability of the system under node or link removal \cite{newman2003structure}, as well as on dynamical processes taking place on the network \cite{Lodato2007}. In order to study such questions systematically, the ability to generate an ensemble of networks with a precise set of topological features, but no others, is crucial. Growth models are generally not suited for this task since a network formation process often introduces invariable correlations between network features that are difficult to disentangle. For example, the Barab\'asi-Albert model is capable of generating networks with a broad degree distribution, but at the same time introduces degree-degree correlations. Further, growth models are generally very difficult to characterize in terms of their statistical properties. In contrast, generative probabilistic models which parameterize an ensemble of networks via an explicit expression for the probability distribution over adjacency matrices can facilitate such analysis. The present work introduces a new class of probabilistic generative models. 

Good generative probabilistic models of networks should combine three characteristics: First, every aspect of network structure that is not explicitly specified through the parameterization is maximally random. Second, they should allow for unbiased estimation of parameters from data. If parameters are estimated from data, these data are typical for the ensemble thus parameterized. Third, they should be easy to specify and parameters should be simple to learn and interpret. Exponential random graph models (ERGMs), \ie those that specify a Boltzmann distribution over the set of all adjacency matrices of given size, meet all of these criteria \cite{Besag1974,Strauss1986,Holland1981,Wasserman1996}. They are maximum entropy, mean unbiased, and parameters can be learned consistently via maximum likelihood estimators or Monte Carlo Markov Chain (MCMC) methods \cite{Snijders2002,Handcock2008,Hunter2008}.

Generally, pairwise relations between nodes, so-called dyads, are considered the fundamental building blocks of complex networks and hence, also the fundamental unit when modeling a network, regardless whether by growth or probabilistic models. 
Erd\"{o}s-Rényi (ER) graphs, the configuration model \cite{Molloy1995,MOLLOY}, stochastic block models \cite{Holland1983,Nowicki2001,Snijders1997,Wang1987} and degree corrected block models \cite{Karrer2010} all fall into the class of dyadic models. The basic assumption underlying dyadic models is that dyads are \emph{conditionally independent} given the model's parameters. 

\begin{figure}
	\centering
		\includegraphics[width=0.49\textwidth]{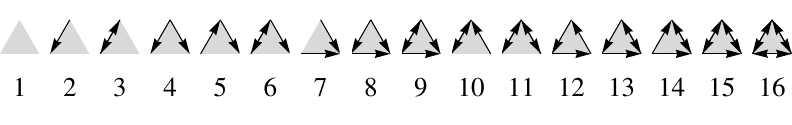}
	\caption{All 16 possible non-isomorphic triadic subgraphs (subgraph patterns) in directed networks.}
	\label{fig:triadPatterns}
\end{figure}

However, the assumption of dyadic independence as a general paradigm of network modeling seems questionable. For example, in a social context, the idea that the relation of Alice and Bob be independent from the relation of Alice and Charlie seems to go against experience, especially if the relation is of romantic type. Similarly, triadic closure, or the large clustering coefficient observed in many networks, hints at a dependence between the connections in a network. Generalizing these ideas, during the last decade the systematic study of third and fourth order sub-network structure captured high attention \cite{Milo2002,Milo2004,Sporns2004,Shen-Orr2002,Albert2004}. Apart from node permutations, there are 16 distinct triad patterns in directed unweighted networks as shown in Fig.~\ref{fig:triadPatterns}. It was found that certain patterns of three-node subgraphs occur significantly more frequent than expected in an ensemble of networks generated by shuffling the connections of the original network under the constraint of preserving the nodes' in- and out-degrees. 

Sampling from an ensemble of randomized networks yields an average occurrence $\left\langle N_{\text{rand},i}\right\rangle$ and a standard deviation $\sigma_{\text{rand},i}$ for each triad pattern $i$ shown in Fig.~\ref{fig:triadPatterns}. Over- and underrepresentation of pattern $i$ is quantified by through a Z-score
\begin{equation}
	Z_{i} = \frac{N_{\text{original},i} - \left\langle N_{\text{rand},i}\right\rangle}{\sigma_{\text{rand},i}}.
	\label{eq:Zscore}
\end{equation}
Notice that Z-scores are evaluated by counting the subgraph patterns over all $\binom{N}{3}$ possible triads. Every network can be assigned a vector $\vec{Z}$ whose components comprise the Z-scores of all possible triad patterns. Significant patterns are referred to as 'motifs' \cite{Milo2002}. It is common to consider only the Z-scores of the triad patterns in which all three nodes are attached to an edge. Further, one commonly refers to the normalized Z-vector as the 'significance profile', $\vec{\text{SP}} = \vec{Z} / \sqrt{ \sum_{i=4}^{16} Z_i^2}$. This normalization makes systems of different sizes comparable \cite{Milo2002}. Many real-world systems have been examined with respect to their triadic Z-scores and significance profiles \cite{Milo2002,Milo2004,Eom2006,Kaluza2010,Sakata2005} and it was suggested that they can be grouped into so-called 'super-families' \cite{Milo2004}.

Surprisingly, to date, no general model exists that can fully explain or model the triad significance profiles observed in many real world networks. The present work suggests a generative probabilistic model capable of describing a wide range of significance profiles.  

A number of growth models exist which are capable of reproducing certain parts of the motif statistics, in particular, the fraction of closed triangles by explicitly formulating 'triadic closure' processes. Starting from an initially unclustered network, one searches for edges with a common neighbor and then connects them successively to form triangles \cite{Newman2009,Jin2001,Holme2001,Klemm2002,Bansal2008,AngelesSerrano2005}. Yet, the calculation of their properties is limited to numerical approaches \cite{Newman2009}.

Further, specifying generative models has proven difficult. Using the Strauss model \cite{Strauss1986}, specified by a Hamiltonian with two fields, one acting on individual links, the other one acting on triads of links, it is possible to generate systems with - on average - predefined link and triad appearance. However, Park and Newman could show that the average does not describe the properties of a typical system generated by the model. In fact, there is a large degenerate phase in which most instances of networks tend to be either fully connected or empty \cite{Park2005}.

Another alternative suggested by Newman generates networks in which both the number of single links, $s_i$, of every node $i$, as well as the number of triads, $t_i$, it participates in are specified initially \cite{Newman2009}. The model yields networks, drawn uniformly at random from the set of all possible matchings of 'stubs' and 'corners'. With this generalization of random-graph models, it is possible to compute analytically component sizes, the existence and size of a giant component, and percolation properties. The model yields an unbiased ensemble of networks with clustering. However, attempting to specify the probabilities for all possible three-node subgraphs simultaneously poses a problem.

Alternatively, it has been noted early on that latent variables might offer an explanation for the observed motif distributions within the framework of dyadic independence models. The randomization employed in the calculation of the Z-scores ignores all mesoscopic structure, possibly present in the system. Thus, parts of the over- and underrepresentations of certain motifs, compared to the randomized versions, may stem from such structure \cite{Reichardt2011,Beber, Fretter2012}. E.g.~ some features of the significance profile of the neural network of \textit{C.~elegans} could successfully be explained by means of latent class structure, while accounting for both properties on the individual node level and on the group level \cite{Reichardt2011}. In~\cite{Beber} the authors show that strong modular structure leads to a strong overrepresentation of subgraph patterns comprised of closed triads. The abundance of triad motifs is apparently strongly related to mesoscopic network structure or, in other words, comparison of a network with block structure to a null model which does not account for such groups may result in Z-scores which are more than less artifacts of the mesoscopic structure. 
Yet, mesoscopic block models alone are not sufficient to explain all observed motifs.

In general, when trying to reproduce triad structures, models formulated in terms of dyads face the difficulty that each dyad influences an extensive number of triads. On the other hand, directly modeling all triad structures is impossible, as not all local triad configurations may be specified independently from each other. Yet, the Z-score statistics are obtained by considering every individual triadic subgraph pattern.
 
In the following section we will suggest a model which is based on triads rather than dyads which actually \textit{can} be specified independently from each other, so-called Steiner Triple Systems (STSs). Starting from the framework of Steiner Triple Systems, it will be possible to define a whole class of triadic exponential random graph models. In this paper we discuss the most basic of such models, one which assumes the same probability distribution of triadic subgraph configurations on all Steiner Triples (STs). This can be considered the triadic analog to ER graphs on dyadic models. We will investigate how a distribution on the STs affects the corresponding triad significance profiles. With this work, we will be able to investigate correlations in the abundance of triad patterns which occur solely for statistical reasons. Moreover, we provide for a class of generative models which are capable of modeling structure of higher than dyadic order. We aim to design ensembles of networks with pre-defined Z-score profiles. In section \ref{sec:STS} we will introduce the concept of STSs, subsequently in section \ref{sec:model} we will define the triadic random graph model, a generative model based on STs. Finally, in section \ref{sec:results} we will present results for the latter. In particular, we will show that triadic random graphs are capable of generating networks with non-vanishing Z-scores. Furthermore, we will investigate correlations in the appearance of triadic subgraph patterns and discuss their implications for the functional interpretation of motif significance profiles. Finally, we will calculate the degree distribution of triadic random graphs analytically.


\section{Steiner Triple Systems} \label{sec:STS}

We will now define the terminology used throughout the remainder of this article: A \textit{dyad} is a set of two nodes. An \textit{edge}, or interchangeably a \textit{link}, describes the presence of a dyadic connection, \ie a connection between two nodes; it can be uni- or bidirectional. A \textit{triad} is a set of three nodes. A \textit{triangle} denotes three mutually interconnected nodes. A \textit{subgraph} is a part of a network which considers only a subset of all nodes, including their mutual connections. A \textit{subgraph configuration} is a specification of the connections in a subgraph, while accounting for node identities; e.g. dyad configuration $A \rightarrow B$ is distinct from dyad configuration $B \leftarrow A$. \textit{Subgraph patterns} are sets of nodes including their relations without accounting for node identities, \ie isomorphic subgraph configurations are mapped to the same subgraph pattern; e.g. dyad pattern $A \rightarrow B$ is the same as pattern $A \leftarrow B$. A(n) \textit{(anti-)motif} is a subgraph pattern which is significantly over-(under-) represented, as compared to some null model.

\begin{figure}
	\centering
	\includegraphics[width=0.2\textwidth]{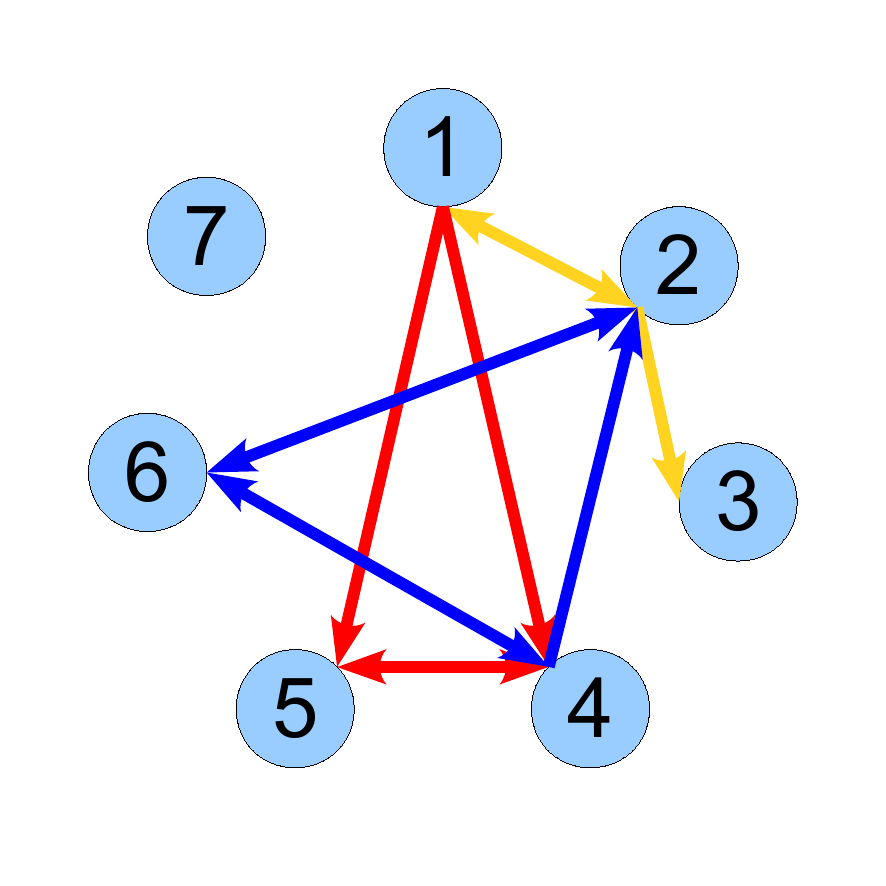}
	\caption{(Color online) Only few triad configurations can be specified independently of each other: e.g. a specification of the triads $\left(1,2,3\right)$, $\left(1,4,5\right)$, and $\left(2,4,6\right)$ fully determines the configuration of $\left(1,2,4\right)$.}
	\label{fig:triadConflict}
\end{figure}

In a network of $N$ nodes there are $T = \binom{N}{3}$ distinct triads. Yet, it is not possible to specify all their triadic-subgraph configurations independently of each other; e.g. consider the network in Fig.~\ref{fig:triadConflict}. Suppose we set the relations in the three-node subgraph of nodes 1, 2, and 3, denoted as $\left(1,2,3\right)$, such that they adopt pattern $\motVI$. Further, we specify the triads $\left(1,4,5\right)$ and $\left(4,6,2\right)$ such that they assume patterns $\motXIV$ and $\motXV$, respectively. Then, with the choices for the discussed three triads in Fig.~\ref{fig:triadConflict}, the subgraph of $\left(4,1,2\right)$ is already determined to take the pattern $\motXIII$ implicitly. This is because $\left(4,1,2\right)$ contains dyadic relations which have already been assigned in the other three triads.

Since there are only $E = \binom{N}{2}$ dyads in a network and every triad comprises three dyadic relations, there is an upper bound to the number of triads which are dyad-disjoint and therefore can be set without over-determining the system:
\begin{equation}
	\text{\# of dyad-disjoint triads} \leq \frac{E}{3} = \frac{N \, \left(N-1\right)}{6} \ll T
	\label{eq:numberDisjointTriads}
\end{equation}

Networks for which the upper bound is exactly met can be partitioned into triads such that every pair of nodes in the system is part of \textit{exactly one} of them. Such systems are called Steiner Triple Systems (STS) \footnote{Steiner triple systems are a special case of the more general $t\left(v,k,\lambda\right)$ or $S_{\lambda}\left(t,k,v\right)$ designs, where $v$ denotes the number of points and $k$ denotes the cardinality of the blocks (three for triangles). For any set $T$ of $t$ points, there are exactly $\lambda$ blocks incident with all points in $T$. Thus, Steiner triple systems are $2\left(v,3,1\right)$ or $S_{1}\left(2,3,v\right)$. For more details see e.g. \cite{VanLint1992}.}. STSs consisting of $N$ vertices are called Steiner Triple Systems of order $N$, or STS($N$). There are two necessary and sufficient requirements for the existence of an STS($N$):
\begin{eqnarray}
	N  \, \text{mod} \, 2 &= 1\label{eq:STS_constr1}\\
	N\left(N-1\right) \, \text{mod} \, 3 &= 0 \label{eq:STS_constr2}
\end{eqnarray}
For a detailed discussion see e.g. \cite[page 277ff]{Hall1986} or \cite[page 205ff]{VanLint1992}. The problem was originally solved by Kirkman in 1847 \cite{Kirkman1847}.

From Eq. \eqref{eq:STS_constr1} and Eq. \eqref{eq:STS_constr2} we can conclude that, by approximation, systems of arbitrary size can be decomposed into Steiner Triples. All one has to do is either add up to three 'dummy' nodes to the system, or to ignore up to three nodes including their relations.

\begin{figure}
	\centering
	\includegraphics[width=0.45\textwidth]{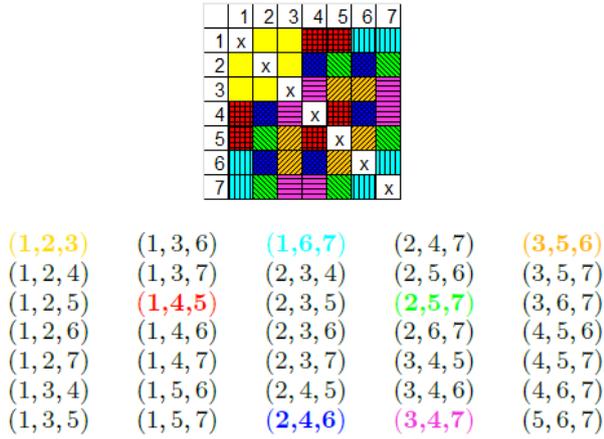}
	\caption{(Color online) Schematic presentation of a Steiner Triple System of order seven. The Steiner Triples are set to be (1,2,3), (1,4,5), (1,6,7), (2,4,6), (2,5,7), (3,4,7), and (3,5,6), as indicated by the colors of the matrix elements. Every matrix element is assigned to exactly one Steiner Triple.}
	\label{fig:STS7matrix}
\end{figure}

To fix ideas, Fig.~\ref{fig:STS7matrix} shows the partition of a STS of order 7 into STs. Due to the small amount of vertices it is possible to derive the STS deductively: Without loss of generality, we start with node 1. Since 1 is part of six dyads (one with every remaining node) it has to be part of three Steiner Triples. The first one shall be (1,2,3) (color coded in yellow in the matrix representation in Fig.~\ref{fig:STS7matrix}), the second one (1,4,5) (red, checkered), and the third one (1,6,7) (cyan, vertical lines). Now each dyadic relation 1 participates in is covered by exactly one Steiner Triple. We continue with the dyads of node 2: those with nodes 1 and 3 are already contained in $(1,2,3)$. 4 and 5 are already part of ST $(1,4,5)$ and therefore need to be assigned to distinct ST. We choose 6 to be in the ST with 2 and 4 (blue, diagonally checkered), and thus we have also specified ST $(1,5,7)$ (green, diagonal lines). Continuing with node 3, the dyads with nodes 4, 5, 6, and 7 need to be assigned to ST. 4 is already assigned to STs with 5 and 6. Thus, the two remaining ST are $(3,4,7)$ (magenta, horizontal lines) and $(3,5,6)$ (orange, diagonal lines).

From the $\binom{7}{3} = 35$ possible triads of a network of order seven only $E/3 = 7 \cdot 6 / 6 = 7$ triads can be specified independently from each other.


Also in Fig.~\ref{fig:STS7matrix}, all triads for a network of seven nodes are displayed. A possible choice of an STS is highlighted with colors corresponding to the matrix representation above. Of course, still most triad configurations will be specified implicitly. However, a STS provides for a \textit{maximum set} of triples which can be specified independently of each other.

A detailed proof that Eq. \eqref{eq:STS_constr1} and Eq. \eqref{eq:STS_constr2} are indeed sufficient for the existence of an STS can be found in \cite{VanLint1992}.

Of course, for larger system sizes it is not practical to construct STSs the way described above. However, larger STS can be constructed by merging smaller ones. For the STS(7), the partition described above is unique, apart from relabeling nodes. For STSs of higher order, there are multiple non-isomorphic ways to partition the nodes into Steiner Triples. STSs provide us with sets of triads which can actually be configured without overdetermining dyadic relations. They thus, can be considered a basis to express an adjacency matrix.

In order to account for substructures of higher than dyadic order, our goal is now to define a model based on triadic rather than dyadic entities. Since Steiner Triple Systems assign every dyadic relation, \ie every pair of nodes, to exactly one triad, the specification of the configurations of all Steiner Triples is equivalent to specifying an adjacency matrix $\boldsymbol{A}$. To convince oneself that a formulation of a network in terms of Steiner Triples is equivalent to a formulation in terms of dyads, consider a directed unweighted graph with $N$ vertices. There are $\binom{N}{2}$ dyads. Each dyad $\left(i,j\right)$ may adopt four distinct configurations. Thus, in total there are $4^{\binom{N}{2}} = 2^{2 \binom{N}{2}}$ possible states of the system, \ie distinct adjacency matrices. On the other hand, there are $\binom{N}{2}/3$ distinct Steiner Triples. Each of those triads may assume $2^6 = 64$ distinct configurations (every of the six unidirectional links in the triad may be present or absent). Therefore, again we obtain $64^{\binom{N}{2}/3} = 2^{6\binom{N}{2}/3} = 2^{2 \binom{N}{2}}$ possible states. The argument for undirected graphs is analogous.


\section{Model} \label{sec:model}

Let us recall that dyadic ERGMs assume that the likelihoods for the presence of two edges are conditionally independent of each other. Further, let the matrix $\boldsymbol{D}$ with components $D_{ij} \in \left\{0,1\right\}$ denote the random variables corresponding to the entries of the adjacency matrix $\boldsymbol{A}$. Then, the independence assumption implies for the likelihood of observing an adjacency matrix $\boldsymbol{A}$
\begin{equation}
 \begin{split}
	\mathcal{P}\left(\textbf{D} = \textbf{A} \, | \, \vec{\theta}\right) 
		&= \prod_{i=1}^{N-1} \prod_{j=i+1}
		 \mathcal{P}\left(D_{ij} = A_{ij} , D_{ji} = A_{ji} | \vec{\theta}\right)\\
		&= \prod_{i=1}^{N-1} \prod_{j=i+1}^{N} \mathcal{P}\left(\vec{D}_{(i,j)} = \vec{A}_{(i,j)} | \vec{\theta}\right)
 \end{split}
 \label{eq:dyadicModel}
\end{equation}
where $\vec{\theta}$ includes all parameters of the model. The vector notation on the right hand side accounts for the fact that in directed unweighted networks, there are four possible dyadic relations: $A_{ij} = 0 \wedge A_{ji} = 0$, $A_{ij} = 0 \wedge A_{ji} = 1$, $A_{ij} = 1 \wedge A_{ji} = 0$, and $A_{ij} = 1 \wedge A_{ji} = 1$. They can be combined in a four dimensional indicator vector $\vec{A}_{(i,j)}$ with all components being zero, except for one being one.

We will now employ the concept of Steiner Triple Systems to define the triadic analog to Eq. \eqref{eq:dyadicModel}. Now, instead of assuming the likelihoods of \textit{dyads} to be conditionally independent of each other, we suppose the likelihoods for the configurations on \textit{Steiner Triples} to be conditionally independent. With this assumption, the likelihood of observing an adjacency matrix $\boldsymbol{A}$ factorizes as follows:
\begin{equation}
 \begin{split}
	\mathcal{P}\left(\textbf{D} = \textbf{A} \, | \, \vec{\theta}\right) 
		&= \prod_{\sigma = 1}^{N(N-1)/6} \mathcal{P}\left(\vec{D}_\sigma = \vec{A}_\sigma | \vec{\theta}\right)\\
		&= \prod_{\sigma = 1}^{N(N-1)/6} \vec{\mathcal{P}}\left(\vec{D}_\sigma | \vec{\theta}\right) \cdot \vec{A}_{\sigma}
 \end{split}
 \label{eq:triadicModel}
\end{equation}
where $\sigma$ denotes the Steiner Triples of an STS($N$), $\vec{D}_\sigma$ is an indicator variable for the configuration of Steiner Triple $\sigma$, and $\vec{A}_\sigma$ is a value of this variable. Analogously to Eq. \eqref{eq:dyadicModel}, for each of the vectors exactly one component is unity, while all others are zero, which is equivalent to the fact that a triad cannot be in multiple configurations at the same time. For undirected networks it is $\vec{D}_\sigma \in \left\{0,1\right\}^{8}$, for directed ones it is $\vec{D}_\sigma \in \left\{0,1\right\}^{64}$. Accordingly, it is $\vec{\mathcal{P}}\left(\vec{D}_\sigma | \vec{\theta}\right) \in \left[0,1\right]^8$ or $\left[0,1\right]^{64}$, respectively with the sums of the elements normalized to one. By defining Eq. \eqref{eq:triadicModel} we make the assumption that the likelihoods of Steiner Triple configurations factorize, \ie they are conditionally independent of each other.

\begin{figure}
	\centering
		\includegraphics[width=0.12\textwidth]{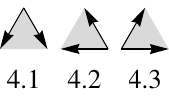}
	\caption{The three isomorphic configurations belonging to pattern 4.}
	\label{fig:triadConfigs4Num}
\end{figure}

For unweighted graphs, Eq. \eqref{eq:triadicModel} describes the most general formulation of models based on conditionally independent STs. We will now further investigate the properties of a particular realization of this class of models. The simplest such model has the same likelihood distribution for the triad configurations on all Steiner Triples, $\vec{\mathcal{P}}\left(\vec{D}_\sigma | \vec{\theta}\right) = \vec{\mathcal{P}}\left(\vec{D} | \vec{\theta}\right)$. Since all nodes are treated equally in this model and a priori, the presence of any link is equally likely, global network architecture is not modeled. This allows us to disentangle the influence of global properties from the impact of local patterns on the overall network structure and to test the hypothesis whether triadic patterns serve as the building blocks of complex networks. The model can be regarded the triadic analog to dyadic ER graphs, in which the likelihood for the existence of an edge is the same for all dyads. We will refer to them as \textit{triadic random graphs}. Since the ordering of the nodes in a Steiner Triple is arbitrary there is no need to distinguish between isomorphic triad configurations. E.g. the likelihoods of the three configurations of subgraph 4, shown in Fig.~\ref{fig:triadConfigs4Num}, will be the same. Thus, the triadic random graphs have 16 parameters each of them indicating the probability of a Steiner Triple to assume one of the subgraphs shown in Fig.~\ref{fig:triadPatterns}. Of course, their values need to sum up to unity.

Given the parameters, the probability distribution of each Steiner Triple is given by:
\begin{widetext}
\begin{equation}
 \begin{split}
	\vec{\mathcal{P}}\left(\vec{D} | \vec{\theta}\right) &= \boldsymbol{M} \, 	\vec{\mathcal{P}}\\&= \boldsymbol{M} \, 	\big( p(\motI) , p(\motII) , p(\motIII) , p(\motIV), p(\motV), p(\motVI), p(\motVII) , p(\motVIII) , p(\motIX) , p(\motX) , p(\motXI) , p(\motXII) ,  p(\motXIII) , p(\motXIV) , p(\motXV) , p(\motXVI) \big)^T
 \end{split}
	\label{eq:motifMapping}
\end{equation}
\end{widetext}
The matrix $\boldsymbol{M}$ maps each of the 16 non-isomorphic subgraph patterns in Fig.~\ref{fig:triadPatterns} to their corresponding isomorphic configurations with equal probability, \ie the sums of its columns are normalized to one. Here, the only parameters $\vec{\theta}$ of the model are the entries of the vector $\vec{\mathcal{P}}$.

Eq. \eqref{eq:triadicModel} and Eq. \eqref{eq:motifMapping} describe the triadic random graph model, in which the configuration for each Steiner Triple is drawn - conditionally independent from other Steiner Triples - from the same probability distribution over the 16 subgraphs shown in Fig.~\ref{fig:triadPatterns}.

If (unidirectional) links are set purely at random with probability $p$ as it is the case in ER graphs, the probabilities for the triadic subgraph patterns are:
\begin{equation}
	\begin{alignedat}{3}
		p_{\motI}^{ER} &= (1 - p)^6\\
		p_{\motII}^{ER} &= 6 \, p \, (1 - p)^5\\
		p_{\motIII}^{ER} &= p_{\motIV}^{ER} = p_{\motVII}^{ER} = 3 \, p^2 \, (1 - p)^4 \,\, , \,\, p_{\motV}^{ER} &= 6 \, p^2 \, (1 - p)^4 \\
		p_{\motVI}^{ER} &= p_{\motVIII}^{ER} = p_{\motX}^{ER} = 6 \, p^3 \, (1 - p)^3  \,\, , \,\, p_{\motXII}^{ER} &= 2 \, p^3 \, (1 - p)^3 \\
		p_{\motIX}^{ER} &= p_{\motXI}^{ER} = p_{\motXIV}^{ER} = 3 \, p^4 \, (1 - p)^2  \,\, , \,\, p_{\motXIII}^{ER} &= 6 \, p^4 \, (1 - p)^2  \\
		p_{\motXV}^{ER} &= 6 \, p^5 \, (1 - p)\\
		p_{\motXVI}^{ER} &= p^6\\
	\end{alignedat}
	\label{eq:motifsERdirected}
\end{equation}
The triadic random graph model allows us to deviate from this probability distribution. Therefore, we can enhance or suppress certain substructures as compared to ER graphs.


\begin{figure*}
	\centering
		\includegraphics[width=0.99\textwidth]{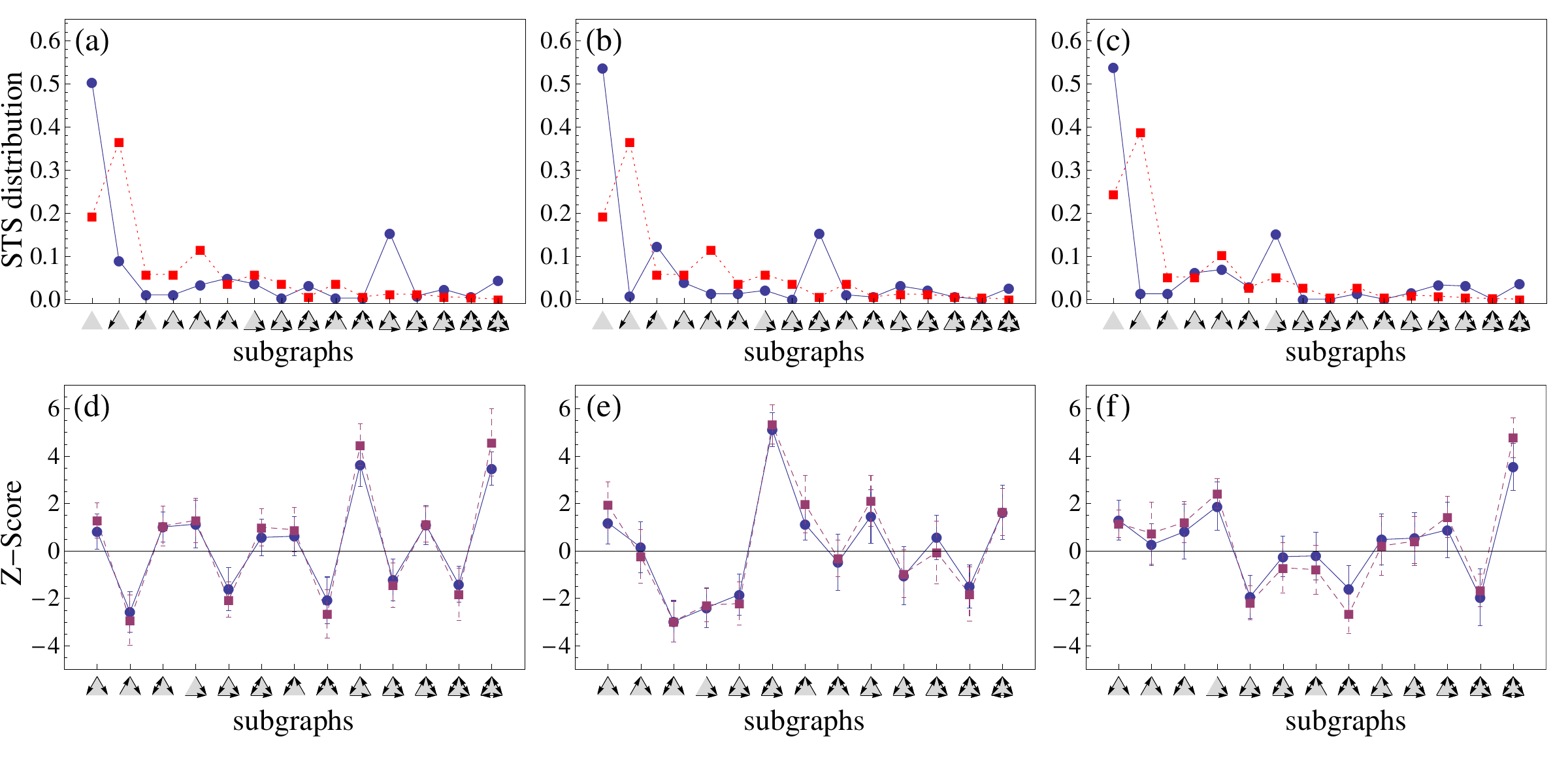}
	\caption{(Color online) \textbf{Top:} Distribution $\vec{\mathcal{P}}$ of triad configurations for the Steiner Triples (blue circles) and expected distribution of triad configurations for ER graphs with the same link density (red squares). \textbf{Bottom:} Z-scores obtained from networks sampled from the distributions above for systems of size $N=49$ (blue circles) and $N=63$ (violet squares), averaged over 15 sample networks.}
	\label{fig:ZscoreProfiles}
\end{figure*}

\begin{figure*}
	\centering
		\includegraphics[width=0.96\textwidth]{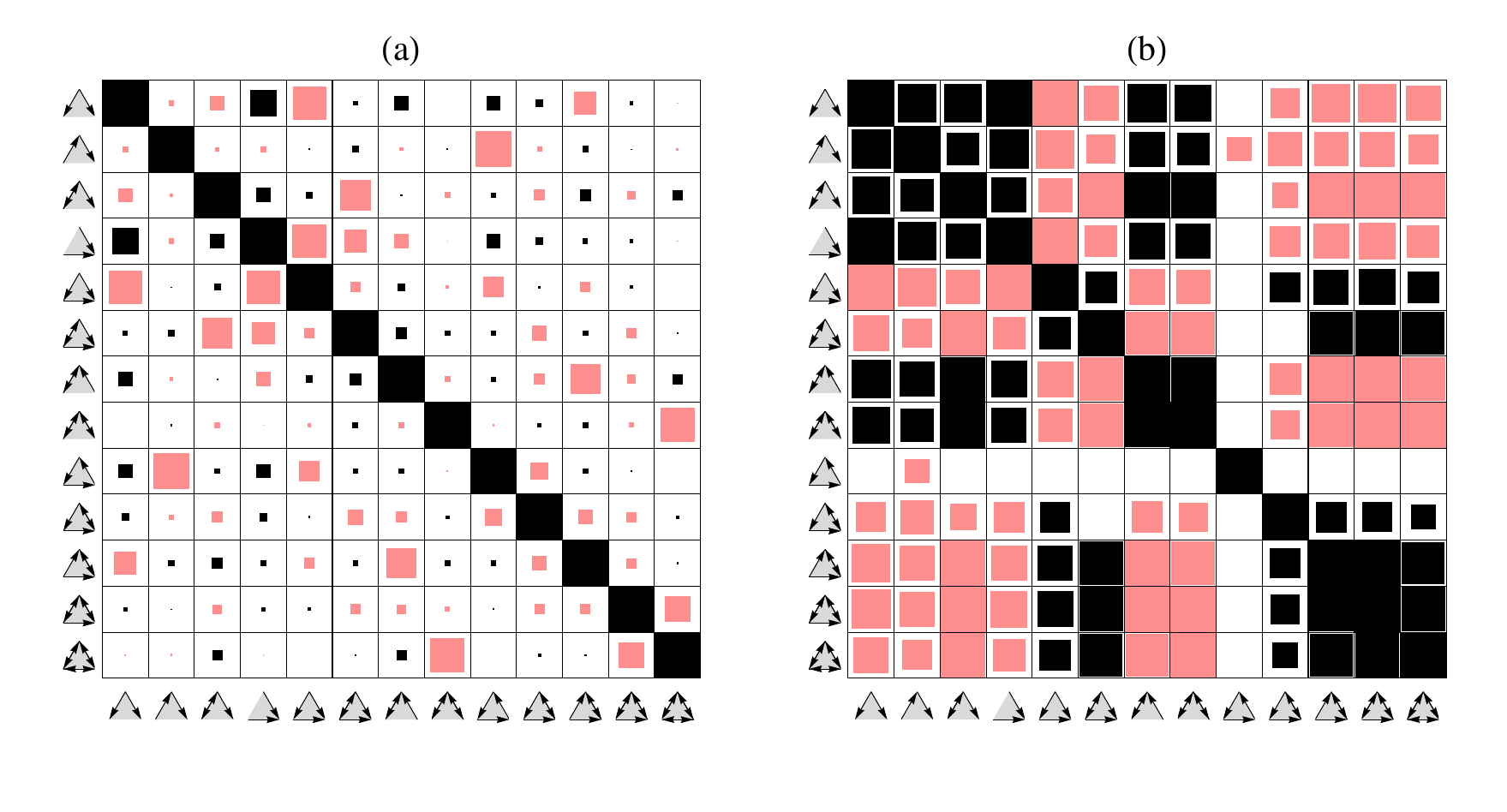}
		\caption{(Color online) \textbf{(a)}~Z-score cross correlations in $2\times10^4$ randomly sampled distributions on Steiner Triple Systems. \textbf{(b)}~Correlations obtained from real data sets (Table \ref{tab:realData}). The length of the squares indicates the magnitude (0 to 1). Black and red shading corresponds to positive and negative values, respectively. Shown are significant entries at a level of 5\%.}
		\label{fig:correlationComp}
\end{figure*}

\section{Results} \label{sec:results}

\subsection{Z-score profiles}

In order to examine the impact of the triad distribution for the STS on the Z-score profile of the total network we did extensive samplings on the 16-dimensional simplex defined by the probability distributions \eqref{eq:motifMapping}. Samplings were performed for both systems of size 49 and 63. For the computation of the Z-score profiles we used the mfinder software (version 1.2)~\cite{AlonWebpage} and averaged the Z-score for each vector $\vec{\mathcal{P}}$ over multiple samples. It shall be noticed that the software only considers those triad configurations which have all three nodes attached to at least one edge. Thus, there are no Z-scores for the subgraphs 1 ($\motI$), 2 ($\motII$), and 3 ($\motIII$) of Fig.~\ref{fig:triadPatterns}. Yet, of course it is necessary to account for them in the input distributions for the STS.

Fig.~\ref{fig:ZscoreProfiles} displays exemplary results obtained from the sampling. Plots a, b, and c show the distributions imposed on the STS (blue circles). This distribution already determines the expectation value for the link-density of the network. E.g. suppose 60\% of the Steiner Triples adopt pattern $\motIV$ (which has two of the six possible links being set)  and 40\% adopt $\motXV$ (five of the six links being set) then the expected density will be $p = (0.6 \cdot 2 + 0.4 \cdot 5) / 6 \approx 53\%$. For comparison we also plot the distribution one would expect on the STs for a dyadic ER graph with the same link density as given by Eq.~\eqref{eq:motifsERdirected} (red squares, dashed line in Fig.~\ref{fig:ZscoreProfiles}).

Plots d, e, and f in Fig.~\ref{fig:ZscoreProfiles} show the Z-score profiles obtained from the input distributions above for networks of size 49 (blue circles) and 63 (violet squares). Displayed are the mean values averaged over 15 samples for each distribution. For systems with no higher order structure, such as ER graphs, all Z-scores are expected to vanish. However, for the triadic random graph model, we observe Z-scores with magnitudes larger than five, implying that certain motifs appear five standard deviations more frequently than expected for the randomized ensemble. Thus, triadic random graphs are capable of modeling structure of higher than dyadic order. It shall be emphasized that this higher order structure does not stem from mesoscopic group structure; all Steiner Triples, and therefore all nodes, have the same parameters. In accordance with the literature~\cite{Milo2002} a larger system size results in a larger magnitude of the Z-scores. However, the shape of the Z-score profiles is size independent.

\begin{table*}
	\centering
		\begin{tabular}{c || c | c | c | c | c | c | c | c | c | c | c | c | c}
			Data set	& $\motIV$ & $\motV$ & $\motVI$ & $\motVII$ & $\motVIII$ & $\motIX$ & $\motX$ & $\motXI$ & $\motXII$ & $\motXIII$ & $\motXIV$ & $\motXV$ & $\motXVI$ \\
			\hline
			\hline
			C. Elegans \cite{NewmanData,Watts1998,White1986} &	-16.5 &	-6.29 &	-24.23 &	-11.99 &	12.43 &	24.48 &	-27.02 &	-16.3 &	-5.02 &	2.59 &	27.29 &	13.15 &	9.64\\
			Political blogs \cite{NewmanData,Adamic2005}	& -76.09 &	-51.28 &	-49.36 &	-58.19 &	55.26 &	40.28 &	-54.07 &	-31.17 &	-2.32 &	2.97 &	47.19 &	27.05 &	24.82\\
			E. Coli (v. 1.1) \cite{AlonWebpage, Mangan2003} &-12.23&-12.23&0&-12.23&12.23&0&0&0&0&0&0&0&0\\
			English book \cite{AlonWebpage,Milo2004} &26.09&13.58&14.35&22.8&-22.52&-10&24.67&13.51&-1.39&-6.59&-21.84&-13.58&-5.53\\
			French book \cite{AlonWebpage,Milo2004} &31.51&26.31&13.4&31.52&-29.1&-10.16&16.17&12.03&-11.5&-12.34&-15.07&-12.33&-4.72\\
			Japanese book \cite{AlonWebpage,Milo2004} &15.01&12.05&13.43&14.97&-14.39&-7.94&12.13&9.27&-4.76&-9.92&-7.4&-8.3&-3.07\\
			Spanish book \cite{AlonWebpage,Milo2004} &26.58&27.5&13.57&23.77&-22.3&-4.16&29.35&12.39&-13.22&-19.82&-25.22&-10.99&-7.57\\
			leader2Inter \cite{AlonWebpage,Milo2004} &-2.25&-1.2&-2.58&-1.22&0.81&1.33&-3.24&-4.5&0.38&1.15&2.31&1.8&3.53\\
			prisonInter \cite{AlonWebpage,Milo2004} &-6.06&-3.71&-10.14&-9.06&4.31&7.84&-8.26&-13.83&0.4&1.99&5.42&7.49&11.93\\
			Electr. circ. (s208) \cite{AlonWebpage} &1.63&-9.57&0&1.63&-1.63&0&0&0&11.01&0&0&0&0\\
			Electr. circ. (s420) \cite{AlonWebpage} &1.61&-17.21&0&1.61&-1.61&0&0&0&20.74&0&0&0&0\\
			S. Cerevisiae \cite{AlonWebpage,Costanzo2001} &-13.73&-13.52&-0.96&-13.66&13.6&-0.35&-5.91&0&-0.17&9.9&3.94&0&0\\
		\end{tabular}
	\caption{Z-scores observed in real-world data sets.}
	\label{tab:realData}
\end{table*}

\subsection{Z-score correlations}

For the interpretation of triad significance profiles observed in real networks it is important to be aware of correlations between the Z-scores of pairs of triad patterns, which inherently already arise solely for statistical reasons.

We did extensive uniform sampling of the 16-dimensional simplex spanned by the parameter space of the triadic random graph model \eqref{eq:motifMapping}. In fact, we sampled more than $2 \times 10^4$ distinct distributions. For each of the distributions, we generated five network instances and we evaluated the average Z-score profiles. Using the latter, we can evaluate cross correlations between pairs of Z-scores over the input distributions sampled. For two patterns, $i$ and $j$, it is
\begin{equation}
	C_{Z_i, Z_j} = \frac{\left\langle Z_i \, Z_j \right\rangle - \left\langle Z_i\right\rangle \left\langle Z_j\right\rangle}{\sigma_{Z_i} \, \sigma_{Z_j}}.
	\label{eq:crosscorrelation}
\end{equation}
The averages are taken over all sampled STS distributions considered for the evaluation of the correlation matrix. The statistical significance of the correlation is tested by means of a t-test.

Fig.~\ref{fig:correlationComp} (a) shows the correlation matrix between pairs of Z-scores when sampling randomly. Considered are significant correlations at a level of 5\%. The side lengths of the squares indicate the magnitudes of the correlation coefficients between the corresponding subgraphs. Positive values are colored in black, negative ones in red. The magnitude (zero to one) is proportional to the length of the squares. One can clearly see that certain Z-scores are strongly anti-correlated with each other while others are positively correlated. To keep track of the impact of the link-density on potential correlations, the distributions are grouped in bins of width 0.05. We evaluated separate correlation matrices for each of the link-density ranges. It turns out that correlations and anti-correlations occur consistently between the same sets of triad patterns for all link-densities sampled.

\begin{table}
	\centering
		\begin{tabular}{c || c | c | c}
			Rank	& patterns & random samples & real data  \\
			\hline
			\hline
			1 &	$\motV$ , $\motXII$ & -0.780487 &	-0.527285 \\
			2 &	$\motXI$ , $\motXVI$ &	-0.74238 &	-0.977566 \\
			3 &	$\motVII$ , $\motVIII$ &	-0.732722 &	-0.998263 \\
			4 &	$\motIV$ , $\motVIII$ &	-0.730034 &	-0.989075 \\
			5 &	$\motVI$ , $\motIX$ &	-0.661864 &	-0.985936 \\
			6 &	$\motX$ , $\motXIV$ & -0.661757 &	-0.996671 \\
			7 &	$\motIV$ , $\motVII$ & 0.578477 &	 0.990511\\
			8 &	$\motXV$ , $\motXVI$ & -0.562423 &	0.958279 \\
			9 &	$\motVII$ , $\motIX$ & -0.49549 &	-0.703186 \\
			10 &	$\motIV$ , $\motXIV$ & -0.488278 &	-0.835307 \\
		\end{tabular}
	\caption{Top 10 (anti-)correlations between subgraph patterns found in the synthetic random samples, as well as the corresponding correlations observed in real-world data sets.}
	\label{tab:correlations}
\end{table}

In order to distinguish between Z-scores which actually describe characteristics of the networks from purely statistical artifacts we also investigated Z-score correlations over various real-world networks. Fig.~\ref{fig:correlationComp} (b) shows the correlation matrix obtained from the 16 real-world data sets shown in  Table \ref{tab:realData}. We observe that the most pronounced correlations found in the ensemble of triadic random graphs also appear in the real data sets. The attribution of functional significance to single (anti-)motifs is therefore difficult. Table \ref{tab:correlations} displays the ten strongest cross correlations between pairs of triadic subgraph patterns which were found in our random samples of the triadic random graph ensemble together with the correlation coefficients found in the real data for the respective pairs of triad patterns. Apparently, nine of the top ten (anti-) correlations of the statistical data are also found in the real systems. However, not all entries of correlation matrix obtained from the triadic random graphs are reflected in Fig.~\ref{fig:correlationComp} b): e.g. patterns $\motXV$ and $\motXVI$ are anti-correlated in the random ensemble, while being strongly positively correlated in the real data. This gives rise to the conjecture that this correlation captures valuable information about the systems' structure. Contrary, e.g. the correlation between patterns $\motIV$ and $\motVII$ seems to stem from statistical roots. 

Investigations of correlations in the appearance of subgraph motifs have been done before by Ginoza and Mugler \cite{Ginoza2010}. Yet, their work focuses on correlations within the randomization process of single networks. They consider motifs in two particular networks, namely for the transcriptional regulatory networks of \textit{E. coli} and \textit{S. cerevisiae}. One of their key results is that the abundances of patterns $\motIV$, $\motV$, and $\motVII$ are strongly mutually correlated, while being anti-correlated with pattern $\motVIII$ in both systems. Moreover, they found correlations between patterns $\motVI$, $\motX$, $\motXIII$, and $\motXIV$ for the \textit{S. cerevisiae} network. Our approach however considers correlations which appear over multiple network instances and is therefore complementary to the one in \cite{Ginoza2010}. Again, Fig.~\ref{fig:correlationComp}~(a), displays our observed correlations between subgraph patterns which occur solely for statistical reasons. In accordance with Ginoza et al. we find strong correlations between patterns $\motIV$ and $\motVII$, as well as strong anti-correlation of them with $\motVIII$. However, the former are hardly correlated with pattern $\motV$ (in fact, the correlation coefficient is even slightly negative). Although, doubtlessly, in most real networks there is a strong mutual (anti-)correlation in the abundance of subgraphs $\motIV$, $\motV$, $\motVII$, and $\motVIII$, our results indicate that they do not necessarily follow for statistical reasons and thus may be of relevance for the performance of the systems' function. Furthermore, in addition to the findings of Ginoza et al., we also observe strong anticorrelations between $\motXI$ and $\motXVI$, between $\motV$ and $\motXII$, between $\motVII$ and $\motIX$, and between $\motVIII$ and $\motIX$.

\subsection{Degree distributions}

An important characteristic of complex networks is their degree distribution.

In dyadic ER graphs, the node degrees are expected to be Poissonian distributed:
\begin{equation}
	\mathcal{P}\left(k = \kappa \right) = e^{-\left\langle k\right\rangle} \frac{\left\langle k\right\rangle^\kappa}{\kappa!}
	\label{eq:Poissonian}
\end{equation}
This holds for both in and out-degrees.

To derive the expected in-degree distribution for triadic random graphs, consider an arbitrary node $i$. It is part of $(N-1)/2$ Steiner Triples. Now let $s_i$ be a random variable indicating the number of $i$'s Steiner Triples in which a single edge is directed towards it. Further, be $d_i$ the random variable indicating the number of its Steiner Triples with two links directed towards it. From the probabilities in Eq. \eqref{eq:motifMapping} we can directly infer the probabilities for a single ST to contribute to $s_i$ and $d_i$, respectively:
\begin{widetext}
\begin{equation}
	\begin{split}
		p\left(s_i\right) =& \frac{1}{3} \left[p\left(\motII\right) + p\left(\motVIII\right) + p\left(\motX\right) + p\left(\motXV\right)\right] + \frac{2}{3} \left[p\left(\motIII\right) + p\left(\motIV\right) + p\left(\motV\right) + p\left(\motIX\right) + p\left(\motXI\right) + p\left(\motXIII\right)\right] + p\left(\motVI\right) + p\left(\motXII\right)\\
		p\left(d_i\right) =& \frac{1}{3} \left[p\left(\motVII\right) + p\left(\motVIII\right) + p\left(\motIX\right) + p\left(\motX\right) + p\left(\motXI\right) + p\left(\motXIII\right)\right] + \frac{2}{3} \left[p\left(\motXIV\right) + p\left(\motXV\right)\right] + p\left(\motXVI\right)\\
	\end{split}
	\label{eq:sdProbabilities}
\end{equation}
\end{widetext}
Since the model parameters are the same for all nodes, the expectation values for $s$ and $d$ will also be the same for all $i$:
\begin{equation}
	\begin{split}
		\left\langle s\right\rangle &= \left\langle s_i\right\rangle = \frac{N-1}{2} \, p\left(s_i\right)\\
		\left\langle d\right\rangle &= \left\langle d_i\right\rangle = \frac{N-1}{2} \, p\left(d_i\right)
	\end{split}
	\label{eq:sdExpectation}
\end{equation}

\begin{figure}
	\centering
	\includegraphics[width=0.49\textwidth]{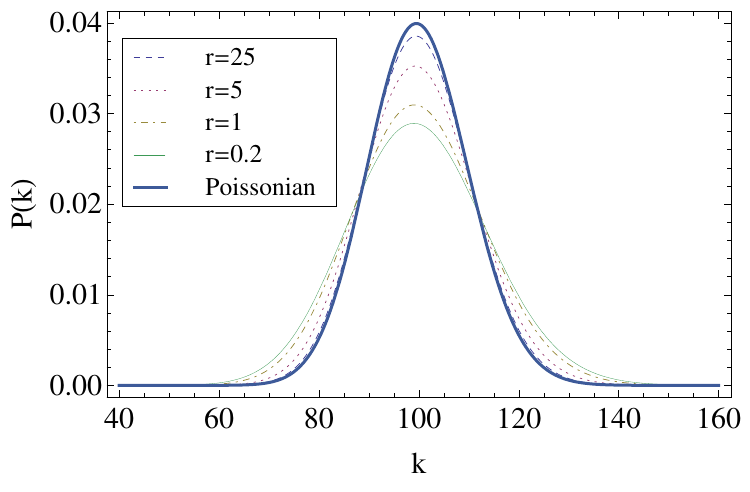}
	\caption{(Color online) Degree distributions for mean degree $\left\langle k\right\rangle = 100$ and various ratios $r = \left\langle s\right\rangle / \left\langle d\right\rangle$.}
	\label{fig:DegDist}
\end{figure}

Each of the $(N-1)/2$ Steiner Triples of node $i$ has either no, one, or two edges directed towards it. Therefore, the joint probability distribution of $s_i$ and $d_i$ is given by the multinomial:
\begin{widetext}
\begin{equation}
	\begin{split}
		p\begin{pmatrix} s_i = n_{s} \\ d_i = n_{d} \end{pmatrix} &= \begin{pmatrix} \frac{N-1}{2} \\ n_{s} \,,\, n_{d} \,,\, \frac{N-1}{2}-n_{s} -n_{d} \end{pmatrix} \, p\left(s_i\right)^{n_{s}} \, p\left(d_i\right)^{n_{d}} \, \left(1- p\left(s_i\right) - p\left(d_i\right)\right)^{\frac{N-1}{2} - n_{s} - n_{d}}\\
								&= \begin{pmatrix} \frac{N-1}{2} \\ n_{s} \,,\, n_{d} \,,\, \frac{N-1}{2}-n_{s} -n_{d} \end{pmatrix} \, \left(\frac{2}{N-1}\right)^{n_{s} + n_{d}} \, \left\langle s\right\rangle^{n_{s}} \, \left\langle d\right\rangle^n_{d} \, \left(1- \frac{2 \, \left(\left\langle s\right\rangle + \left\langle d\right\rangle\right)}{N-1}\right)^{\frac{N-1}{2} - n_{s} - n_{d}}
	\end{split}
	\label{eq:sdDistribution}
\end{equation}
\end{widetext}
For the second equality we used Eq. \eqref{eq:sdExpectation}. For large sparse systems and $\left\langle s\right\rangle , \left\langle d\right\rangle \ll N$ we find
\begin{widetext}
\begin{equation}
	\begin{split}
		\lim_{N \to \infty} p\begin{pmatrix} s_i = n_{s} \\ d_i = n_{d} \end{pmatrix} =& \lim_{N \to \infty} \frac{\left(\frac{N-1}{2}\right)!}{\left(\frac{N-1}{2} - n_{s} - n_{d}\right)!} \, \left(\frac{2}{N-1}\right)^{n_{s} + n_{d}} \, \frac{\left\langle s\right\rangle^{n_{s}}}{n_{s}!} \, \frac{\left\langle d\right\rangle^{n_{d}}}{n_{d}!}\\&
								 \times \, \underbrace{\left(1- \frac{\left\langle s\right\rangle + \left\langle d\right\rangle}{\frac{N-1}{2}}\right)^{\frac{N-1}{2}}}_{\to e^{-\left\langle s\right\rangle -\left\langle d\right\rangle}} \, \underbrace{\left(1- \frac{\left\langle s\right\rangle + \left\langle d\right\rangle}{\frac{N-1}{2}}\right)^{-\left(n_{s} + n_{d}\right)}}_{\to 1}\\
								=& \lim_{N \to \infty} \underbrace{\frac{\frac{N-1}{2} \times \frac{N-3}{2} \times ... \times \frac{N-1}{2} - n_{s} - n_{d} +1}{\left(\frac{N-1}{2}\right)^{n_{s} + n_{d}}}}_{\to 1} \, \frac{\left\langle s\right\rangle^{n_{s}}}{n_{s}!} \, \frac{\left\langle d\right\rangle^{n_{d}}}{n_{d}!} \, e^{-\left\langle s\right\rangle -\left\langle d\right\rangle}\\
								=& \, \frac{\left\langle s\right\rangle^{n_{s}}}{n_{s}!} \, \frac{\left\langle d\right\rangle^{n_{d}}}{n_{d}!} \, e^{-\left\langle s\right\rangle -\left\langle d\right\rangle}.
	\end{split}
	\label{eq:sdDistributionLim}
\end{equation}
\end{widetext}
The in-degree of node $i$ is
\begin{equation}
	k^{\text{in}}_{i} = s_i + 2 d_i.
	\label{eq:ksd}
\end{equation}
The probability distribution for node $i$ to have in-degree $\kappa$ is thus
\begin{equation}
	\begin{split}
		p\left(k^{\text{in}} = \kappa \right) =& \sum_{n_{s} = 0}^{\frac{N-1}{2}} \, \sum_{n_{d} = 0}^{\frac{N-1}{2}} \, p\begin{pmatrix} s_i = n_{s} \\ d_i = n_{d} \end{pmatrix} \boldsymbol{\delta}_{\kappa,n_{s} + 2n_{d}}\\
									=& e^{-\left\langle s\right\rangle -\left\langle d\right\rangle} \, \sum_{n_{d} = 0}^{\frac{\kappa}{2}} \, \frac{\left\langle s\right\rangle^{\kappa-2\, n_{d}}}{\left(\kappa-2n_{d}\right)!} \, \frac{\left\langle d\right\rangle^{n_{d}}}{n_{d}!}
	\end{split}
	\label{eq:degDistribution}
\end{equation}
where $\boldsymbol{\delta}$ is the Kronecker delta ($\boldsymbol{\delta}_{i,j} = 1$ if $i = j$, 0 otherwise). In the limit $\left\langle d\right\rangle \to 0$ the distribution is Poissonian. With $\left\langle d\right\rangle$ approaching $\frac{1}{2} \left\langle k^{in}\right\rangle$, the distribution 
becomes broader, implying larger deviations from $\left\langle k\right\rangle$. Fig.~\ref{fig:DegDist} shows distributions of Eq.~\eqref{eq:degDistribution} with fixed $\left\langle k\right\rangle = \left\langle s\right\rangle + 2 \left\langle d\right\rangle = 100$ for various ratios of $r = \left\langle s\right\rangle / \left\langle d\right\rangle$ together with the corresponding Poissonian.

The out-degree distribution can be derived analogously. In this case, only the probabilites for the triads with a single out-going edge, $p\left(s_i^\text{out}\right)$, and for two out-going edges, $p\left(d_i^\text{out}\right)$, need to be adjusted accordingly.
\begin{widetext}
\begin{equation}
	\begin{split}
		p\left(s_i^\text{out}\right) =& \frac{1}{3} \left[p\left(\motII\right) + p\left(\motVI\right) + p\left(\motVIII\right) + p\left(\motXV\right)\right]
											+ \frac{2}{3} \left[p\left(\motIII\right) + p\left(\motV\right) + p\left(\motVII\right) + p\left(\motXI\right) + p\left(\motXIII\right) + p\left(\motXIV\right)\right] + p\left(\motX\right) + p\left(\motXII\right)\\
		p\left(d_i^\text{out}\right) =& \frac{1}{3} \left[p\left(\motIV\right) + p\left(\motVI\right) + p\left(\motVIII\right) + p\left(\motXI\right) + p\left(\motXIII\right) + p\left(\motXIV\right)\right]
											+ \frac{2}{3} \left[p\left(\motIX\right) + p\left(\motXV\right)\right] + p\left(\motXVI\right)\\
	\end{split}
	\label{eq:sdProbabilitiesOut}
\end{equation}
\end{widetext}

\begin{figure}
	\centering
		\includegraphics[width=0.484\textwidth]{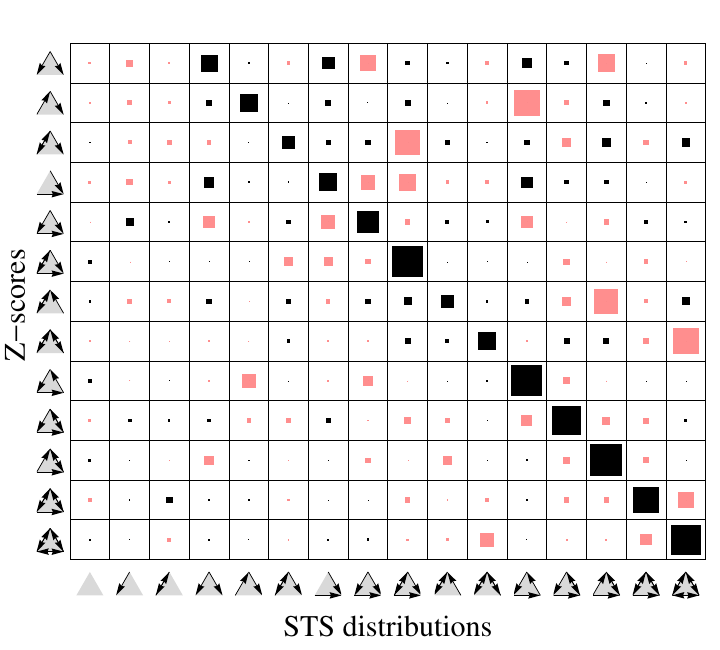}
		\caption{(Color online) Correlation matrix between triad configurations on ST and the resulting Z-score profiles obtained from 5000 configurations. The length of the squares indicates the magnitude (0 to 1). Black and red shading corresponds to positive and negative values, respectively.}
		\label{fig:STSZcorrelation}
\end{figure}

\subsection{Design of significance profiles}

To achieve the goal of designing networks with certain triad significance profiles, it is important to understand the relationship between the distribution of triad configurations on Steiner Triples and the Z-scores obtained from their ensembles. Therefore, we also investigated the cross correlations between the ST configurations and the obtained corresponding Z-scores:
\begin{equation}
	\widetilde{C}_{\mathcal{P}_i, Z_j} = \frac{\left\langle \mathcal{P}_i \, Z_j \right\rangle - \left\langle \mathcal{P}_i\right\rangle \left\langle Z_j\right\rangle}{\sigma_{\mathcal{P}_i} \, \sigma_{Z_j}}.
	\label{eq:crosscorrelation2}
\end{equation}
Results are presented in Fig.~\ref{fig:STSZcorrelation}. Of course, there is a strong correlation between the imposed triad patterns on the Steiner Triples and the Z-scores of these patterns. However, as for the Z-score-Z-score cross correlations, again we observe strong anti-correlations between certain patterns. As before, the observations are valid for all examined link densities. Correlations between the input distributions on the STS and the obtained over-all Z-score profiles can be helpful in designing systems with pre-defined significance profiles.

\begin{figure}
	\centering
		\includegraphics[width=0.493\textwidth]{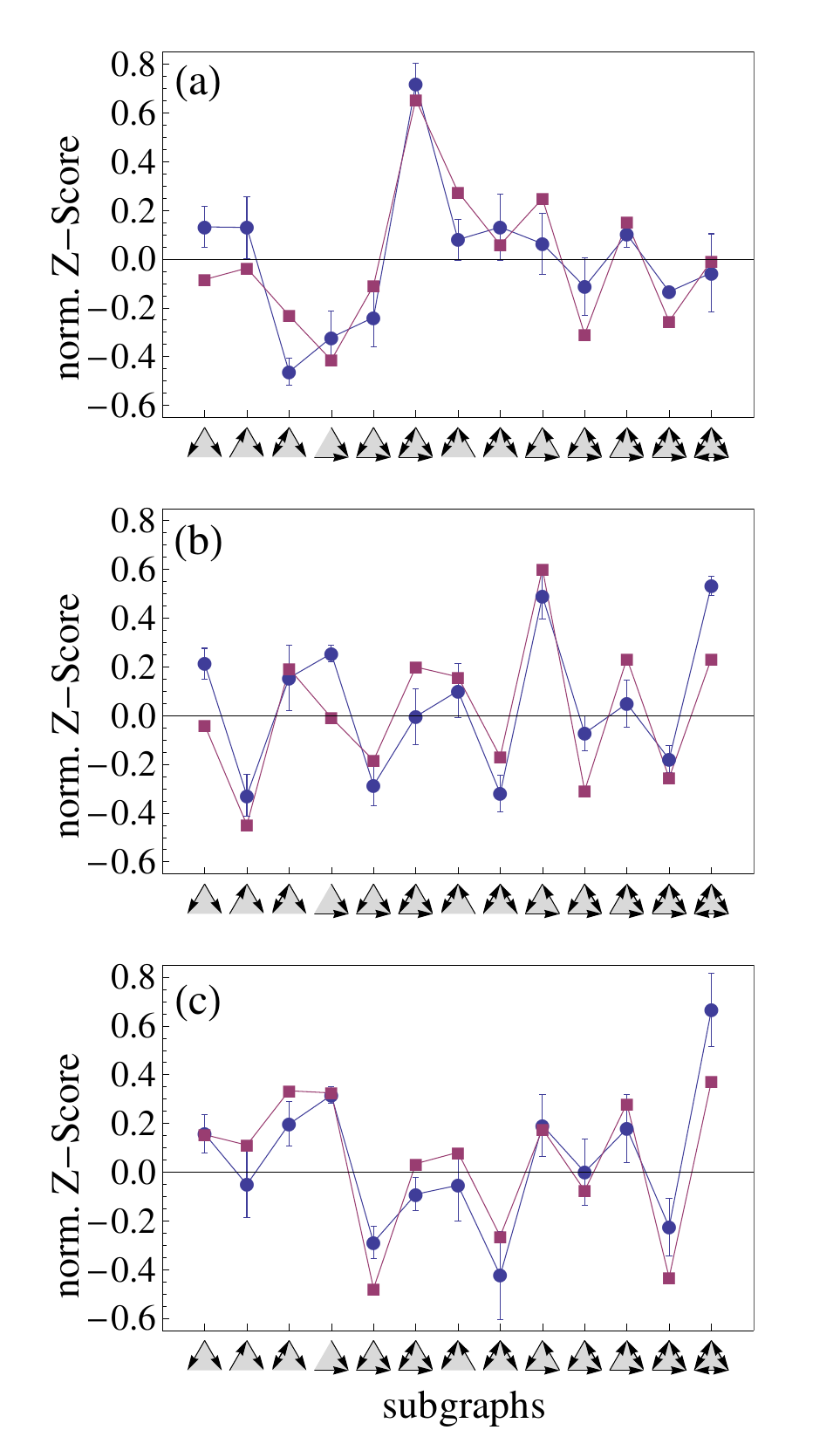}
		\caption{(Color online) Significance profiles corresponding to the Z-scores in Fig.~\ref{fig:ZscoreProfiles} for systems of size 49 (blue circles). The violet squares indicate the prediction obtained from the input distribution $\vec{\mathcal{P}}$ by assuming $\vec{\text{SP}} \propto \boldsymbol{\widetilde{C}} \, \vec{\mathcal{P}}$.}
		\label{fig:STSSPprediction}
\end{figure}

\begin{figure}
	\centering
		\includegraphics[width=0.43\textwidth]{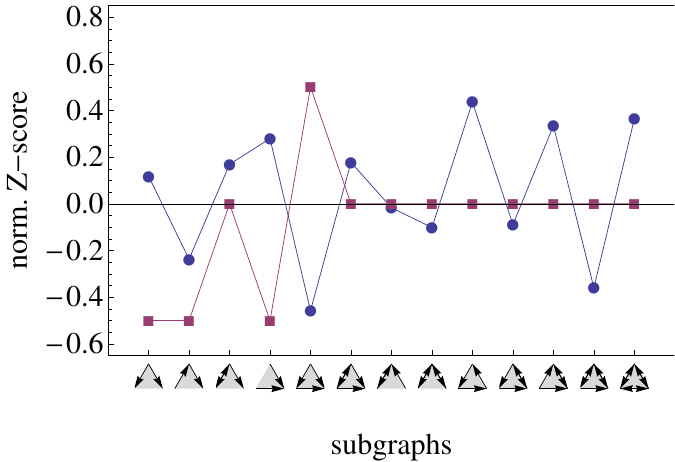}
		\caption{(Color online) Attempt to model the significance profile indicated by the violet squares. However, networks with their parameterization obtained from Eq. \eqref{eq:SPdesign} yield very different significance profiles.}
		\label{fig:STSSPdesign}
\end{figure}

For a simplistic approach, we assume a linear relation between the input distribution $\vec{\mathcal{P}}$ and the significance profile, conveyed by the correlation matrix $\boldsymbol{\widetilde{C}}$ (Fig.~\ref{fig:STSZcorrelation}),
\begin{equation}
	\vec{\text{SP}} \propto \boldsymbol{\widetilde{C}} \, \vec{\mathcal{P}}.
	\label{eq:SPpred}
\end{equation}
In order to design systems with pre-defined significance profiles, it is necessary to map the latter to a corresponding input distribution, which can be realized by means of the pseudo inverse matrix $\boldsymbol{\widetilde{C}}^{-1}$.
\begin{equation}
	\vec{\mathcal{P}} \propto \boldsymbol{\widetilde{C}}^{-1} \, \vec{\text{SP}}
	\label{eq:SPdesign}
\end{equation}
Fig.~\ref{fig:STSSPprediction} shows the significance profiles corresponding to the Z-scores in Fig.~\ref{fig:ZscoreProfiles} together with the prediction obtained from Eq.~\eqref{eq:SPpred}. The predictions agree very well with the actually observed profiles. However, attempts to model arbitrary significance profile with the linear relation Eq.~\eqref{eq:SPdesign} will not succeed in all cases as shown in Fig.~\ref{fig:STSSPdesign}.

This may be for various reasons. On the one hand the relationship between $\vec{\mathcal{P}}$ and the significance profile is certainly not entirely linear. Secondly, not all significance profiles are necessarily realizable, e.g. think of a SP with all patterns being overrepresented. Furthermore, the triadic random graph model describes the most simplistic model based on STS, which, e.g. does not account for individual node properties. This is also reflected in the fact that the degree distributions of triadic random graphs are close to a Poissonian. A formulation of more specific models based on STS may overcome these shortcomings. Still, these first steps open the way to efficiently generate networks in which certain motifs are over- or underrepresented and thus enable systematic investigations of the functional significance of these motifs.


\section{Conclusions}

Over the last decade the over- and underrepresentation of particular sub-network patterns has attracted high attention. This led to the hypothesis that, instead of links, they serve as the building blocks of network structures~\cite{Milo2002}. The fact that only a small portion of triad configurations can actually be specified independently poses a challenge to the formulation of generative models which account for higher order substructures. Based on sets of pair-disjoint triads, so called Steiner Triple Systems, we have introduced a novel class of generative models. The simplest realization of such models assumes the same probability distribution over the possible triad patterns for all Steiner Triples in the system. We referred to them as triadic random graph models. By extensive samplings we proved that, in contrast to ER graphs, even this most simplistic model is capable of inducing non-vanishing Z-scores. Furthermore, we discovered inevitable correlations between triad patterns with respect to their abundance. These occur solely for statistical reasons. This dependence in the appearance of subgraph patterns should be taken into account when attributing functional relevance to network motifs in real systems. Moreover, we unveiled correlations between the probability distributions on the Steiner Triples and the observed Z-score profiles over the whole network. These are helpful for designing ensembles of networks with pre-defined significance profiles which can facilitate a systematic study of the effect of motif distributions on network dynamics. Finally, we could also calculate the degree distributions of triadic random graphs analytically. We found it to be similar, yet not identical to a Poissonian. Depending on the input distribution $\vec{\mathcal{P}}$, the degree distribution is broader than a Poissonian.

The triadic random graph model assumes all nodes to be equal and thus can be considered the triadic analog to ER graphs. However, in many real-world systems, individual node properties like the popularity or activity of vertices play a crucial role. Future models based on Steiner Triple Systems which, e.g. aim to predict hitherto undiscovered links may include those parameters in Eq.~\eqref{eq:triadicModel} in order to model the correct degree distribution. In addition, the introduced framework also allows for the definition of models for signed networks, \ie graphs with positive or negative edges which play in important role in the social sciences in the context of structural balance theory \cite{heider1944social,cartwright1956structural} as well as in the biosciences where they are used to model excitatory and inhibitory links in neural or gene-regulation networks~\cite{Li2004}.


\bibliographystyle{apsrev4-1}

\end{document}